\documentclass[11pt]{article}
\newcommand{\Comment}[1]{{}}
\usepackage{amsfonts,amsthm,amsmath,amssymb,slashed}
\usepackage[textwidth = 430 pt, textheight = 630 pt]{geometry}

\Comment{\usepackage{color}
\definecolor{MyDarkBlue}{rgb}{0.15,0.15,0.45}
\usepackage[linktocpage=true]{hyperref}
\hypersetup{
colorlinks=true,
citecolor=MyDarkBlue,
linkcolor=MyDarkBlue,
urlcolor=MyDarkBlue,
pdfauthor={Cristhiam Lopez-Arcos, Horatiu Nastase, Francisco Rojas and Jeff Murugan},
pdftitle={Conductivity in a gravity dual to massive ABJM and the membrane paradigm},
pdfsubject={hep-th}
}

\usepackage[numbers,sort&compress]{natbib}
\usepackage{hypernat}}
\usepackage{graphicx}
\usepackage{cite}

\newcommand\ignore[1]{}
\def\one{{\,\hbox{1\kern-.8mm l}}}

 \newcommand{\pd}{\partial}

\def\a{\alpha}\def\b{\beta}
\def\g{\gamma}
\def\G{\Gamma}\def\e{\epsilon}

\def\m{\mu}

\def\L{\Lambda}

\def\d{\partial}

\newcommand{\Cset}{{\,\,{{{^{_{\pmb{\mid}}}}\kern-.45em{\mathrm C}}}}}

\newcommand{\be}{\begin{equation}}
\newcommand{\bea}{\begin{eqnarray}}

\newcommand{\ee}{\end{equation}}
\newcommand{\eea}{\end{eqnarray}}

\newcommand{\ret}{\nonumber \\}
\newcommand{\nn}{\nonumber}

\begin{document}

\renewcommand{\thefootnote}{\fnsymbol{footnote}}

\makeatletter
\@addtoreset{equation}{section}
\makeatother
\renewcommand{\theequation}{\thesection.\arabic{equation}}

\rightline{}
\rightline{}


\vspace{10pt}


\begin{center}
{\LARGE \bf{\sc Conductivity in the gravity dual to massive ABJM and the membrane paradigm}}
\end{center} 
 \vspace{1truecm}
\thispagestyle{empty} \centerline{
{\large \bf {\sc Cristhiam Lopez-Arcos${}^{a,}$}}\footnote{E-mail address: \Comment{\href{mailto:crismalo@ift.unesp.br}}{\tt crismalo@ift.unesp.br}},
{\large \bf {\sc Horatiu Nastase${}^{a,}$}}\footnote{E-mail address: \Comment{\href{mailto:nastase@ift.unesp.br}}{\tt nastase@ift.unesp.br}}, 
{\large \bf {\sc Francisco Rojas${}^{a,}$}}\footnote{E-mail address: \Comment{\href{mailto:frojasf@ift.unesp.br}}{\tt frojasf@ift.unesp.br}}
{\bf{\sc and}}
{\large \bf {\sc Jeff Murugan${}^{b,}$}}\footnote{E-mail address: \Comment{\href{mailto:jeff@nassp.uct.ac.za}}{\tt jeff@nassp.uct.ac.za}}
                                                           }

\vspace{.5cm}

\centerline{{\it ${}^a$ 
Instituto de F\'{i}sica Te\'{o}rica, UNESP-Universidade Estadual Paulista}} \centerline{{\it 
R. Dr. Bento T. Ferraz 271, Bl. II, Sao Paulo 01140-070, SP, Brazil}}

\centerline{{\it ${}^b$
The Laboratory for Quantum Gravity \& Strings, }} \centerline{{\it
Department of Mathematics and Applied Mathematics, }} \centerline{{\it
University of Cape Town, Private Bag, Rondebosch, 7700, South Africa.}}

\vspace{1truecm}

\thispagestyle{empty}

\centerline{\sc Abstract}

\vspace{.4truecm}

\begin{center}
\begin{minipage}[c]{380pt}
{\noindent In this paper we analyze the effect of the massive deformation of the ABJM model on the calculation of conductivity of the dual theory. We show that some of the difficulties presented by the dual geometry, in particular the construction of black holes therein, can be at least partially circumvented by adopting a membrane paradigm-like computation of the conductivity, which requires us to know just the effect of the deformation on the {\it horizon} of a black hole in AdS${}_{4}$. The deformation at the horizon itself is found by first deforming the flat space near the horizon, and then using the corresponding solution near the horizon as initial conditions for the Einstein's equations. We find the same result, showing an increase in conductivity, using two types of membrane paradigm 
computations.
}
\end{minipage}
\end{center}

\vspace{.5cm}

\setcounter{page}{0}
\setcounter{tocdepth}{2}

\newpage

\renewcommand{\thefootnote}{\arabic{footnote}}
\setcounter{footnote}{0}

\linespread{1.1}
\parskip 4pt


\newcommand{\bean}{\begin{eqnarray*}}
\newcommand{\eean}{\end{eqnarray*}}


\section{Introduction}
\ \ \ \ \ 
The past decade has seen the gauge/gravity correspondence emerge as a radically new computational paradigm in understanding quantum field theories at strong coupling. While the duality between type IIB superstrings on $AdS_{5}\times S^{5}$ and $\mathcal{N}=4$ supersymmetric $SU(N)$ Yang-Mills theory remains the canonical example for such strong coupling computations, the recently discovered duality between type IIA superstrings on $AdS_{4}\times \mathbb{CP}^{3}$ and the so-called ABJM model \cite{Aharony:2008ug}, a $(2+1)-$dimensional $\mathcal{N}=6$, $U(N)\times U(N)$ Chern-Simons-matter theory is quickly gaining ground in popularity. Much of the interest in this version of the duality stems from the utility of the ABJM model in testing various strongly coupled phenomena in planar condensed matter systems. One such application is the relatively poorly understood quantum critical phase that appears at nonzero temperatures around the $T=0$ transition point between 
insulators and superconductors (superfluids). Since this system is effectively described by a conformal field theory (see for instance \cite{Sachdev:2011wg} and references therein), it is a prime candidate for application of the AdS/CFT toolbox.

Expanding on this point, the physics of particles (relevant on the insulator side of the transition) or vortices (relevant on the superconducting phase of the 
transition) suggest different behaviours for the conductivity of the system as a function of frequency, $\sigma(\omega)$. In \cite{Myers:2010pk} it was suggested 
that this situation could perhaps be captured by the gravity dual of the ABJM model by coupling the global $U(1)$ field to the Weyl tensor. Then, depending on the coupling, $\gamma$, one obtains either the particle-like or vortex-like behaviour for $\sigma(\omega)$. However, at nonzero temperatures, the conformal symmetry is broken by the scale $T$ and understanding how this deformation affects quantities like the conductivity is essential. To this end finding a laboratory in which this can be done in a controlled setting that preserves as much of the other symmetries as possible is key.

In this paper we study just such a laboratory, the massive deformation of the ABJM model of \cite{Gomis:2008vc,Terashima:2008sy}, that is known to preserve the full ${\cal N}=6$ supersymmetry. We want to generalize the calculation of \cite{Myers:2010pk} to include the effect of the mass deformation, however we will 
find that we can only calculate the {\it DC conductivity}, $\sigma(0)$. Even in this limited case though, in order 
to calculate this, we need to resort to a membrane paradigm-type calculation, similar to the one in \cite{Iqbal:2008by}, and extend its results to a more general set-up. An important technical point is that, in the absence of the exact solution for the black hole in the massive 
deformation of the gravity dual of ABJM, we need to rely on a perturbative approach to the massive deformation around the horizon of the black hole. We believe that this approach is new, novel and may indeed prove useful in other contexts as well.

The paper itself is organized as follows. In section 2 we review the methods and results of \cite{Myers:2010pk}. In section 3 we show that by applying the Kubo formula, only now for a boundary term at the horizon of the AdS black hole, we obtain exactly the same DC conductivity as \cite{Myers:2010pk}, and then review and generalize another membrane paradigm calculation as defined in \cite{Iqbal:2008by}. In section 4 we perform the massive deformation of the black hole horizon, by first finding the 
zeroth and first order deformations, then using the Einstein's equations to compute the higher order one. We benchmark our procedure against the known AdS black hole with excellent agreement. Section 5 concerns itself with a calculation of the conductivity using the two versions of the membrane paradigm calculation. We conclude the formal part of the paper with a discussion of future directions in section 6. Following this, in a series of appendices, we elaborate on some of the details omitted in the main text. In Appendix A outlines the T-duality calculation from the type IIB to IIA backgrounds, while in Appendix B we show that the presence of a non-vanishing $g_{rt}$ contribution to the metric, which appears due to the massive deformation, doesn't affect the 
formulas for the membrane paradigm calculations with the example of a scalar field. This is then applied, in Appendix C, to the case of the conductivity calculation. Finally, Appendix D, provides the details of the calculation using the Kubo formula at the horizon.

\section{Conductivity for AdS black holes}

To begin, and largely to establish our conventions and notation, let us review quickly the computation of the conductivity associated to a Maxwell field in an $AdS$ black hole background following \cite{Myers:2010pk}. Materials possessing a quantum critical phase, like ultracold ${}^{87}Rb$, exhibit a zero temperature phase transition in terms of the coupling 
$g$ of the system, between a superconducting phase for $g<g_c$ and an insulator phase for $g>g_c$. Modelling the order parameter of the system as a bosonic (scalar) field, the material is typically described by the action 
\be
S=\int d^3x\left[|\d_\tau \phi|^2+v^2|\vec{\nabla}\phi|^2+(g-g_c)|\phi|^2+\frac{n}{2}|\phi|^4\right].
\ee
From this action, it is clear that when $g<g_c$, the order parameter $\langle\phi\rangle\neq 0$, resulting in the usual superfluid properties of the material, while when $g>g_c$ we have that $\langle\phi\rangle=0$, 
giving an insulator phase. Thus $(g=g_c, T=0)$ is a critical point of the system, described by a $(2+1)-$dimensional conformal field theory. To apply the arsenal of the gauge/gravity correspondence to problems like this, it is assumed that this system is dual to some  (quantum) gravitational theory in a one higher dimensional $AdS_4$ space. For example, the $(2+1)-$dimensional ABJM model that is the focus of this article, has a gravity dual theory in $AdS_4\times \mathbb{CP}^3$. 

At nonzero temperatures a quantum critical phase opens up\footnote{See e.g. figure 2 in \cite{Sachdev:2011wg}.} around $g=g_c$ and it is expected that the physics of this quantum critical point should be governed by the zero temperature conformal field theory. 
However, a computation of the frequency-dependent conductivity $\sigma(\omega)$ using a particle/hole description which is generally valid on the insulator side of the phase transition, yields a functional form that {\it decreases to a minimum} before stabilizing. On the other hand, a similar computation using a vortex description, valid on the superfluid side of the phase transition, obtains a conductivity that {\it increases to a maximum} before stabilizing (see for instance figures 4 and 5 in \cite{Sachdev:2011wg}). There seems to be no way to choose which behaviour should be realized using field theory arguments. A 
semi-phenomenological parameter will be introduced in the gravity dual that will govern this choice ($\gamma>0$ for particle-like and $\gamma<0$ for vortex-like).

The problem appears to be better suited to a dual gravity description. To this end, let's consider the gravity dual of the ABJM model reduced to just Einstein-Maxwell gravity in $AdS_4$ (plus possible terms higher order 
in curvature and the field strength tensor $F_{\mu\nu}$). If we consider only terms that vanish on the $AdS_4$ background but that are nonzero in the $AdS_{4}-$black hole background corresponding to a finite temperature field theory, there is a unique contribution to the action: a coupling to the Weyl tensor. With this in mind, the action for the Maxwell field in the gravity dual can be taken to be 
\be
  S_{vector}=\frac{1}{g_4^2}\int d^4x\sqrt{-g}\Big[-\frac{1}{4}F_{\mu\nu}F^{\mu\nu}+\gamma L^2C_{\mu\nu  
  \rho\sigma}F^{\mu\nu}F^{\rho\sigma}\Big],\label{mxaction}
\ee
in a probe approximation for the background. That is, we consider that the AdS black hole background is not modified by the gauge field perturbation, and is given by
\be
  ds^2=\frac{r_0^2}{L^2u^2}\left(-f(u)dt^2+dx^2+dy^2\right)+\frac{L^2du^2}{u^2f(u)},\label{adsbh}
\ee
where
\bea
  f(u)&=&1-u^3,\cr
  r_0&=&\frac{4\pi T L^2}{3}. \label{fulcom}
\eea 
With this in place, the DC conductivity can now be calculated from a membrane-paradigm like calculation starting from a more general form of the action in (\ref{mxaction}),
\be
  S=\int d^4x\sqrt{-g}\left[-\frac{1}{8g_4^2}F_{\mu\nu}X^{\mu\nu\rho\sigma}F_{\rho\sigma}\right].\label{xffaction}
\ee
The action (\ref{mxaction}) corresponds to a choice of
\be
  {X_{\mu\nu}}^{\rho\sigma}=2{\delta_{[\mu\nu]}}^{[\rho\sigma]}-8\gamma L^2{C_{\mu\nu}}^{\rho\sigma}.
\ee
A natural current $j^\mu$ can be defined on the horizon by considering the horizon boundary term associated with (\ref{mxaction}) and varying it with respect to $A_\mu$. Using the equations of motion, we find the conductivity as the proportionality constant in $j^x=\sigma F_{0x}$, giving
\be
\sigma_0\equiv \sigma(\omega=0,k=0)=\frac{1}{g_4^2}\sqrt{-g}\sqrt{-X^{xtxt}X^{xrxr}}\Big|_{r=r_H},\label{conduct}
\ee
which leads to\footnote{The DC conductivity defined here is at $\omega=0$; more generally, we will see shortly that the relevant parameter is 
$w=\omega/(4\pi T)$, hence the result is valid for $\omega/T\ll 1$. This is the quantity that will be of interest for us in the remaining sections of the 
paper.}
\be
\sigma_0=\frac{1}{g_4^2}(1+4\gamma).
\ee
However, this membrane paradigm calculation is somewhat obscure.\footnote{In particular, because it is not obvious that this definition of conductivity
at the horizon is the same as the definition at the boundary.} We will return to another membrane-paradigm like computation in the next section, but 
first, let's note that one can also do the usual AdS/CFT calculation utilising the Kubo formula, 
\be
  \sigma(\omega)=-{\rm Im}\left(\frac{G_{yy}(\vec{q})}{\omega}\right),\label{kubo}
\ee
to calculate the frequency-dependent conductivity.
Here $G_{yy}$ is the 2-point function for the gauge field with $y$ component, $A_y$.
Defining $u\equiv r/r_0$, so that $u=1$ at the horizon and $u=0$ at the boundary, and Fourier transforming gives
\be
  A_\mu(t,x,y,u)=\int \frac{d^3q}{(2\pi)^3}e^{i\vec{q}\cdot\vec{x}}A_\mu(\vec{q},u),
\ee
where $\vec{q}\cdot\vec{x}=-t\omega+q^xx+q^yy$, with $q^\mu=(\omega,q,0)$. The boundary condition at the horizon is 
\be
A_y(\vec{q},u)=(1-u)^bF(\vec{q},u),
\ee
at $u=1$, with $F(\vec{q},u)$ regular there. To compute the conductivity associated to $A_y$, we first calculate the action for $A_y$ on-shell, obtaining the boundary term
\be
  S_{yy}=-\frac{1}{2g_4^2}\int d^3x\left.\left[\sqrt{-g}g^{uu}g^{yy}(1-8\gamma L^2{C_{uy}}^{uy})A_y(u,
  \vec{x})\d_u A_y(u, \vec{x})\right]\right|_{\rm boundary}\label{bdy}.
\ee
Evaluating this expression on the AdS black hole solution, we get
\be
  S_{yy}=-\frac{2\pi T}{3g_4^2}\int d^3x\left.\Big[(1-u^3)(1+4\gamma u^2)A_y(u,\vec{x})\d_u A_y(u,\vec{x})
  \Big]\right|_{\rm boundary}\label{syy}.
\ee
Considering that the boundary is only at $u=0$ (infinity), we obtain
\bea
  S_{yy}&=&-\frac{2\pi T}{3g_4^2}\int d^3x\left.\Big[A_y(u,\vec{x})\d_u A_y(u,\vec{x})\Big]\right|_{u=0}\cr
  &\equiv& \int\frac{d^2\vec{q}}{(2\pi)^3}\frac{1}{2}A_y(-\vec{q})G_{yy}(\vec{q})A_y(\vec{q}),
\eea
where
\be
  G_{yy}(\omega, q=0)=\left.-\frac{4\pi T}{3g_4^2}\frac{\d_u A_y(u,\omega)}{A_y(u,\omega)}
  \right|_{u\rightarrow 0}.
\ee
Substituting this into the Kubo formula then gives
\be
  \sigma=\left.\frac{1}{3g_4^2}{\rm Im}\frac{\d_u A_y}{wA_y}\right|_{u\rightarrow 0},
\ee
where $w\equiv \omega/4\pi T$. Now, the equation of motion for $A_y$ with sources on the boundary can be solved. Indeed, for small $w$ (or, equivalently, small $\omega$) the solution is found to be 
\be
  A_y(u)\simeq (1-u)^{-iw}\left[F_1(u)+wF_2(u)\right],\label{ay}
\ee
where $F_1(u)=C$ (constant) and $F_2'(0)=iC(2+12\gamma)$. This, together with the fact that both $F_1(u)$ 
and $F_2(u)$ are well-behaved at the horizon $u=1$, means that 
\be
\sigma(\omega\rightarrow 0)=\frac{1}{3g_4^2}{\rm Im}\Big[i+\frac{F_2'(0)}{F_1(0)}\Big]=\frac{1}{3g_4^2}[1+(2+12\gamma)]=\frac{1+4\gamma}{g_4^2},\label{}
\ee
exactly the same as the earlier membrane-paradigm calculation.  

In ABJM theory (at $\gamma=0$), we can find the meaning of $g_4$ as follows. The maximally gauged supergravity action in 4 dimensions has a coupling determined
by $L$ and the 4d Newton's constant $\kappa_N^{(4)}$, 
\be
g_4\propto \frac{L^{-1}}{\kappa_N^{(4)-1}}=\frac{\kappa_N^{(4)}}{L}
\ee
where the proportionality factor is a numerical constant. Since by compactifying on $S^7/{\mathbb Z}_k$, with $L_{S^7}=2L$,
\be
\frac{1}{[\kappa_N^{(4)}]^{-1}}=\frac{\Omega_7 (L_{S^7})^7}{[\kappa_N^{(11)}]^2k}=\frac{\Omega_7 g_s^{-3}l_s^{-9}2^7L^7}{k},
\ee
where $\Omega_7$ is the volume of the unit 7-sphere, we obtain
\be
g_4\propto g_s^{3/2}k^{1/2}\left(\frac{l_s}{L}\right)^{9/2}\propto \lambda^{1/4} N^{-1}\label{g4}
\ee
where in the last equality  we wrote $g_4$ in terms of field theory parameters using $L/l_s=5^{5/4}\sqrt{\pi}\lambda^{1/4}$ and $g_s=\lambda^{5/4}/N$.

\section{Membrane paradigm and conductivity from horizon data}\label{sec:membrane}

In order to extend the computation of the conductivity to the mass-deformed ABJM model, we will first need to modify the above membrane-paradigm computation to account for the fact that we are, as yet, unable to construct a black hole directly in the gravitational dual. The modification itself is easy enough to state; in (\ref{bdy}), the boundary term was considered to be at $u=0$ (at infinity), whereas now we want to consider the same boundary term at the horizon $u=1$. This is a bit unusual, since it assumes that the sources for the bulk are {\it at the horizon} and that, strictly speaking, we have to ignore the normal boundary at $u=0$, since otherwise we get a divergent contribution due essentially to the infinite ratio between the contributions from the two boundaries. Let's first benchmark this against the above case of the $AdS_{4}$ black hole to check that we are indeed on the right track before looking for a better justification for it.\\

For the boundary term at $u=1$ only, from (\ref{syy}) we get
\be
S_{yy}=-\frac{2\pi T}{3g_4^2}(1+4\gamma)\int d^3x \left.\Big[3(1-u)A_y(u,\vec{x})\d_u A_y(u,\vec{x})\Big]\right|_{ u\rightarrow 1},
\ee
which leads to
\be
G_{yy}(\omega, q=0)=-\frac{4\pi T}{3g_4^2}(1+4\gamma)\left.\left[3(1-u)\frac{\d_u A_y(u,\omega)}{A_y(u,\omega)}\right]\right|_{u\rightarrow 1}.
\ee
This expression, together with the Kubo formula (\ref{kubo}), evaluated at the horizon the approximation $\d_u A_y/w\simeq i/(1-u)$ yields 
\be
\sigma(\omega\rightarrow 0)=\frac{1}{3g_4^2}(1+4\gamma){\rm Im}\left[3(1-u)\frac{i}{1-u}\right]=\frac{1+4\gamma}{g_4^2}.
\ee
Not only is this the same result as the standard membrane paradigm computation but notice also that
the factors of $1+4\gamma$ and $-3$ enter in a completely different way, so it is highly nontrivial that we get the same answer.\footnote{Also notice that
the temperature has cancelled in the calculation of the DC conductivity, which as already explained, is still valid as long as $\omega/T\ll 1$.}

To better understand this result, let's reconsider the general membrane paradigm analysis carried out in \cite{Iqbal:2008by} where the same type of current at the horizon as in \cite{Myers:2010pk} was considered in the context of the action
\be
S_{em}=-\int_{\Sigma} d^{d+1}x\sqrt{-g}\frac{1}{4g^2_{d+1}(r)}F_{MN}F^{MN}\label{raction}
\ee
where $g_{d+1}(r)$ is an $r$-dependent coupling possibly arising from background fields. The bulk action results in a boundary term that can be cancelled by 
\be
S_{bd}=\int_\Sigma d^dx\sqrt{-\gamma}\left(\frac{j^\mu}{\sqrt{-\gamma}}\right)A_\mu,
\ee
where $j^\mu$ is obtained by varying the action with respect to $\d_r A_\mu$, and $\gamma_{\mu\nu}$ is the induced metric on the stretched horizon $\Sigma$. The same calculation as in the previous section then leads to 
\be
\sigma(\omega\to0)=\frac{1}{g^2_{d+1}(r_0)},\label{dc1}
\ee
which is indeed compatible with (\ref{conduct}) for $d=3$ and ${X_{\mu\nu}}^{\rho\sigma}=2{\delta_{[\mu\nu]}}^{[\rho\sigma]}$. 

The equations of motion of the action (\ref{raction}) in a diagonal background
\be
ds^2=-g_{tt}dt^2+g_{rr}dr^2+g_{ij}dx^idx^j
\ee
then imply 
\be
\d_r j^i=0+{\cal O}(\omega F_{it});\;\;\;\; \d_r F_{it}=0+{\cal O}(\omega j^i)\label{zeroes}
\ee
which in turn mean that the relation 
\be
j^i(r_H)=\left.\frac{1}{g^2_{d+1}}\sqrt{\frac{-g}{g_{rr}g_{tt}}}g^{zz}\right|_{r_H} F_{it}(r_H)\label{jvsF}
\ee
is valid at all $r$, until infinity, with the same coefficient. Therefore the DC conductivity calculated at the horizon equals the one calculated at 
infinity, which is the usual AdS/CFT formula, with the result
\be
\sigma(\omega=0)=\left.\frac{1}{g^2_{d+1}}\sqrt{\frac{-g}{g_{rr}g_{tt}}}g^{zz}\right|_{r_H}
\ee
which matches (\ref{dc1}) for $d=3$. In order to apply these methods to our action in (\ref{mxaction}), we  shall consider the case when we can write, on the background solution,
\be
  S=-\int d^4x \sqrt{-g} \left(\frac{1}{4g_{MN,4}^2(r)}F_{MN}F^{MN}\right).\label{effmaxwell}
\ee
Moreover, for the conductivity associated to $A_y$, we need only the $F_{ry}$ component to be nonzero in which case, following the membrane paradigm calculation, we obtain,
\be\label{conductivity}
\sigma(\omega=0)=\frac{1}{g^2_{ry,4}}
\ee
instead of (\ref{dc1}) so that (\ref{zeroes}) still holds, implying that the membrane paradigm calculation still matches the AdS/CFT result as was indeed verified for pure ABJM theory. We therefore feel confident in using it for the massive deformation. Having said that, one potential issue could be the fact that we will need to deal with a more general metric background of the form
\be\label{metrica}
  ds^2=-g_{tt}(r,t)dt^2 +g_{rr}(r,t) dr^2 +2g_{rt}(r,t)dr\,dt+ g_{yy}(r)(dx^2+dy^2) 
\ee
However, as we show in Appendix B, the off-diagonal terms do not influence the result. Indeed, we will see shortly that the same is true for the calculation of the conductivity. 

We should also point out that an explicit time dependence in the metric, as in \eqref{metrica}, modifies \eqref{zeroes} by replacing the ``0'' on the right hand side of the first equation by time derivatives of the metric 
(see Appendix C for more details). This would imply that we could not use (\ref{zeroes}) as an argument to show that the conductivity defined at the horizon matches the one defined at the boundary. However, this does not necessarily mean they are different, since the {\it integrated} $r-$dependence for $\sigma=j^i/F_{it}$ could still give the same result at the horizon and the boundary. This is as expected from the validity of AdS/CFT, which is, after all, a relation between the boundary field theory (the conductivity being a boundary observable) and bulk gravitational physics, and the membrane paradigm which implies that the this bulk gravitational physics should be describable in terms of just a fictitious membrane at the horizon. Moreover, the time dependence cannot last forever. The metric 
must ``relax'' to a stationary one after some time has passed, as expected from general principles of black hole physics.  Thus although, in principle, there could be an explicit time dependence in the bulk metric, we can assume this dependence will die out way before we ``shake'' the system with the gauge fields $A_{\mu}^A$ we are using as probes.   
 
\section{Massive deformation for ABJM black hole}
As alluded to in the introduction to this article, the ABJM model admits a mass deformation that, while obviously breaking its conformal symmetry, preserves its full $\mathcal{N}=6$ supersymmetry \cite{Gomis:2008vc,Terashima:2008sy}. Unfortunately, the gravity dual to this massive deformation is, to say the least, very complicated (see \cite{Auzzi:2009es,Mohammed:2010eb} for more details), and finding an exact black hole solution in this case seems 
hopeless. What we {\it can} do however, is to consider the deformation of the background in which the M2-branes of the ABJM model live. Recall that the pure ABJM model arises from the near-horizon limit of the back-reacted geometry produced by M2-branes moving in the background $\mathbb{R}^{2,1}\times \mathbb{C}^4/\mathbb{Z}_k$, i.e. 11-dimensional flat space with a $\mathbb{Z}_k$ identification. The background corresponding to the mass-deformed ABJM model is obtained from the maximally supersymmetric pp-wave of type IIB string theory by a sequence of T-dualizing and lifting to M-theory \cite{Mohammed:2010eb,Lambert:2011eg}. To add a black hole to this construction is a notoriously difficult task.

\subsection{First order deformation near the horizon}

Fortunately, as argued earlier, it will suffice for us to consider just a mass-deformation of the horizon of the $AdS_{4}$ black hole. To zeroth order, we can consider the horizon of the 
black hole to be flat, so we can trivially add a pp-wave to it by replacing the flat space with the pp-wave space, in the coordinates around the spherical horizon. The mass deformation is then implemented T-dualizing to obtain a first order solution. To obtain a (back-reacted) solution around the horizon of the black hole with the mass deformation, we will input this first order solution as initial data into the Einstein's equations, and obtain the required metric and Weyl tensor at the horizon. We will show how to carry this out explicitly in the next subsection and, for now, restrict our attention to obtaining the first order solution to be used as an initial data.

To superpose the pp-wave on to the AdS black hole, we must first obtain flat space in usual Minkowski coordinates. To do that, we will take a near-horizon limit and then perform a double Wick rotation. 
We detail these steps as follows. The starting point is the $AdS_4\times S^{7}$ black hole metric in Poincar\'e coordinates
\bea\label{AdS4BH}
ds^2= \frac{r^2}{L^2}\left[-\left(1-\frac{r_0^3}{r^3}\right)dt^2 + d\vec{x}^2\right] + 
L^2\frac{dr^2}{r^2\left(1-{r_0^3 \over r^3}\right)} + L^2 d\Omega_7^2,
\eea
where $\vec{x}=(x_1,x_2)$. Since the near horizon geometry corresponds to $r-r_0\ll r_0,L$, we should also consider $L\,\Omega_7$ large, 
and approximate it with the flat $d\vec{y_7}^2$. Changing the radial coordinate $r$ to $\rho \equiv 2\sqrt{r/r_0-1}$ and expanding in 
powers of $\rho$ yields the near horizon geometry, 
\bea
ds_M^2 \simeq \frac{r_0^2}{L^2}\left[-\frac{3}{4}\rho^2 dt^2+dx_1^2+dx_2^2\right] +  
\frac{L^2}{3} d\rho^2 + d\vec{y}_7^2.
\eea
Since
\bea
g_{tt} = -\frac{3r_0^2}{4L^2}\rho^2, \qquad g_{x_1x_1}=\frac{r_0^2}{L^2,} \qquad 
g_{\rho\rho}=\frac{L^2}{3},
\eea
reducing to string theory on $x_2$ and T-dualizing along $x_1$ yields
\bea
  \tilde{g}_{x_1x_1}&=&\frac{1}{g_{x_1x_1}}=\frac{L^2}{r_0^2},\nonumber\\
  \tilde{g}_{\rho\rho}&=&g_{\rho\rho}=\frac{L^2}{3},\nonumber\\
  \tilde{g}_{tt}&=&g_{tt}=-\frac{3r_0^2\rho^2}{4L^2},\\
  \tilde{g}_{\mu x_1}&=&0,\nonumber\\
  \tilde{\phi}&=&-\frac{1}{2}\ln \left(\frac{r_0^2}{L^2}\right).\nonumber
\eea
The type IIB metric can then be read off as
\bea
  ds_{IIB}^2 &=& - \frac{3r_0^2}{4L^2}\rho^2dt^2 + \frac{L^2}{r_0^2}dx_1^2 
  + \frac{L^2}{3}d\rho^2  + dy_7^2.
\eea
For notational convenience, in what follows, we define the new variables,
\bea
  \label{newvars}
  \tau \equiv \frac{3 r_0}{2L^2}t, \qquad \bar{\rho} \equiv 
  \frac{L}{\sqrt{3}}\rho, \qquad \bar{x}_1 \equiv \frac{L}{r_0}x_1,
\eea
in terms of which,
\bea
  ds_{IIB}^2 &=& - \bar{\rho}^2 d\tau^2 + d\bar{x}_1^2 
  + d\bar{\rho}^2  + dy_7^2.
\eea
Wick rotating to $\tau=i\theta$ yields
\bea
  ds^2 &=& \bar{\rho}^2 d\theta^2 + d\bar{x}_1^2 
  + d\bar{\rho}^2  + dy_7^2.
\eea
Now we Wick rotate back along a Euclidean coordinate by first defining
\bea
  z_1&\equiv&\bar{\rho}\cos \theta = \bar{\rho} \cos(-i\tau) = \bar{\rho} \cosh \tau \\
  z_2&\equiv&\bar{\rho}\sin \theta = \bar{\rho} \sin(-i\tau) = -i \bar{\rho} \sinh \tau.
\eea
and rotating to $z_2=-i\bar{\tau}$, to give the new set of variables
\bea
  z_1&=& \bar{\rho} \cosh \tau, \\
  \bar{\tau}&=&\bar{\rho} \sinh \tau. 
\eea
Near $\tau\sim 0$ we have $z_1\simeq \bar{\rho}$ and $\bar{\tau}\simeq \bar{\rho} \tau$, so that $\bar{\rho} \simeq z_1$ and $\tau\simeq \bar{\tau} / \bar{\rho}$. Thus, expanding in powers of $\tau$ 
\bea
  d\tau^2 &=& \frac{d\bar{\tau}^2}{\bar{\rho}^2}+\mathcal{O}(\tau), \\
  d\bar{\rho}^2 &=& dz_1^2+\mathcal{O}(\tau), 
\eea
yields
\bea
  ds^2 &=& - d\bar{\tau}^2 + d\bar{x}_1^2 
  + dz_1^2  + dy_7^2\label{flat}.
\eea
Now that we have the near horizon metric explicitly written as a flat metric, we simply add the pp-wave to the horizon by replacing (\ref{flat}) with
\bea
ds^2=  - d\bar{\tau}^2 + d\bar{x}_1^2 
- \mu^2\left(z_1^2+\vec{y}_7^2\right)\left(d\bar{\tau}+d\bar{x}_1\right)^2 + dz_1^2  + dy_7^2,
\eea
and turning on an $F_{+z_1\cdots}$, where the lightcone coordinate $x^{+}\equiv\bar{x}_1+\bar{\tau}=\bar{x}_1+\bar{\rho}\sinh\tau$ and $z_1=\bar{\rho}\cosh\tau$, and the ellipses are three of the $\vec{y}$. Note that $\mu$, the mass parameter of the pp-wave added, corresponds in the gauge theory to the mass deformation
parameter of the same name, as was argued in \cite{Mohammed:2010eb,Lambert:2011eg}. In particular, from the identification in eq. 2.28 of \cite{Lambert:2011eg}
with the calculation of the same quantity in gauge theory, we can see that $\mu$ is indeed the same parameter. 
Going back from the ($z_1,\bar{\tau}$) variables to the original ($\bar{\rho},\tau$) ones gives the 
IIB wave solution
\be\label{IIBsolution}
ds^2=  - (d(\bar{\rho} \sinh \tau ))^2 + d\bar{x}_1^2 
- \mu^2\left(\bar{\rho}^2 \cosh^2 \tau+\vec{y}_7^2\right)\left(d(\bar{\rho} \sinh \tau)+d\bar{x}_1\right)^2 + (d(\bar{\rho} \cosh \tau))^2  + dy_7^2.
\ee
In order to arrive at the desired eleven dimensional metric, we now need to T-dualize it to IIA and then lift it to M-theory. The Buscher rules applied to this 
metric are displayed in Appendix \ref{App:Tduality} and, after some algebra, the resulting type IIA metric turns out to be
\bea
ds^2_{IIA}&=&\frac{1}{1-\mu ^2 \left(\bar{\rho} ^2 \cosh^2\tau+y_7^2\right)}
\Big(-\bar{\rho}^2d\tau^2 + d\bar{\rho}^2 + d\bar{x}_1^2\cr
&& - \mu ^2 \left(\bar{\rho} ^2 \cosh^2\tau+y_7^2\right)(d(\bar{\rho}\cosh\tau))^2\Big) 
+ d\vec{y}_7^2.
\eea
Now, returning to the original ($\rho, t, x_1$) variables using \eqref{newvars}, and defining
\bea
H\equiv 1-\mu ^2 \left(\frac{L^2}{3} \rho ^2 \cosh^2\left[\frac{3 t r_0}{2 L^2}\right] + y_7^2\right)
\eea
puts the type IIA metric above into the form
\be
ds^2_{IIA}= H^{-1}\left[-\frac{3r_0^2}{4L^2}\rho^2dt^2 + \frac{L^2}{3}d\rho^2 + 
\frac{L^2}{r_0^2}dx_1^2 - (1-H)\frac{L^2}{3}\left(d\left(\rho\cosh\frac{3r_0t}{2L^2}\right)\right)^2\right]+ d\vec{y}_7^2.
\ee
Finally, lifting to M-theory gives
\bea\label{mmetric}
ds_M^2&=& H^{-2/3}\Bigg[-\frac{3r_0^2}{4L^2}\rho^2dt^2 + \frac{L^2}{3}d\rho^2 + 
\frac{L^2}{r_0^2}dx_1^2 +dx_{11}^2 \ret 
&&- (1-H)\frac{L^2}{3}\left(d\left(\rho\cosh\frac{3r_0t}{2L^2}\right)\right)^2 \Bigg] + H^{1/3} d\vec{y}_7^2.
\eea
This metric corresponds to mass-deforming the black hole solution, though only near the horizon. We also note that (\ref{flat}) only corresponded to the metric near the 
horizon for small $\bar \tau$, or equivalently for small $t$, so we will still need to be careful to use it only in black hole calculations in this region.

Having, through considerable ingenuity, obtained the eleven dimensional metric, in order to apply the procedure for calculating conductivity, we first need to dimensionally reduce it to four dimensions with a cosmological constant. 
The question at hand then is whether we can perform a consistent truncation to four dimensions. The general ansatz for the truncation of the metric on an $n$-sphere $S^n$ down to $d$ dimensions is \cite{Nastase:2000tu} (here $D=d+n$)
\be
  ds_D^2=\Delta^{-\frac{2}{d-2}}ds_d^2+\Delta^\b T_{AB}^{-1}DY^ADY^B
\ee
where
\bea
  \b&=&\frac{2}{n-1}\frac{d-1}{d-2},\nonumber\\
  \Delta^{2-\b n}&=&T^{AB}Y_AY_B,\\
  DY^A&=&dY^A+B^{AB}Y^B.\nonumber
\eea
Here $Y^A(y)$, with $A=0,...,n$, the scalar spherical harmonics on the sphere parametrized by $\vec{y}$ intrinsic coordinates, are also coordinates for the Euclidean embedding of the sphere, i.e.
$Y^AY^A=1$. In our case, $d=4,n=7$ and $B^{AB}=0$, giving
\bea
  ds_{11}^2&=&\Delta^{-1}ds_4^2+\Delta^{1/2}T_{AB}^{-1}dY^AdY^B,\cr
  \Delta&=&(T^{AB}Y_AY_B)^{-2/3}.
\eea
This also matches with another expression found in \cite{Cvetic:1999xx}, for $T_{AB}=X_A\delta_{AB}$, with 
\be
  ds_D^2=(X_AY_A^2)^{\frac{2}{d-1}}ds_d^2+(X_AY_A^2)^{-\frac{d-3}{d-1}}(X_A^{-1}dY_A^2).
\ee
In our case, both this and the former expressions reduce to:
\be
  ds_{11}^2=(X_{A}Y_{A}^2)^{2/3}ds_{4}^2 + (X_{A}Y_{A}^2)^{-1/3}X_{A}^{-1}dY_{A}^2\label{redansatz}
\ee
where $A=0,...,7$. The question is whether the metric \eqref{mmetric} can be described as this kind of consistent truncation to four dimensions on a deformed 7-sphere since we need a metric 
solving a four dimensional action obtained by consistent truncation. It is easy to see that, in general, it can not. However, as we will now demonstrate, it is possible to do so in a special limit. 

The spherical harmonics $Y^A$ are coordinates for the embedding of the sphere in 8-dimensional Euclidean space, i.e. $Y_{A}Y^{A}=L^2$. We will consider a 
small patch of the sphere, namely near the ``North Pole", $Y^0=L$. Then we can consider an intrinsic parametrization of the sphere, i.e. some set of coordinates $\vec{y}$ which match the $\vec{y}_7$ coordinates in our metric. Taking the $y_i$'s to be small means that the spherical harmonic components are 
$\overrightarrow{Y_7}(\vec{y}_7) \simeq \overrightarrow{y_7}$ so that, from the sphere constraint, we have
\be
  Y^{0}=\sqrt{L^2-(\overrightarrow{Y_7})^2}\simeq\sqrt{L^2-(\overrightarrow{y_7})^2}.  
\ee
The factor $\Delta$ in the metric is then
\be
X_{A}Y_{A}^2 = X_0 \left(L^2-(\overrightarrow{y_7})^2\right) + X_7^i ({y}^i_7)^2,
\ee
and, reading off equation from \eqref{mmetric}, this should match with $H^{-1}$. Expanding both sides in powers of $(y_7^i)^2$ and matching coefficients, yields the eight scalars
\bea\label{scalars}
X_0L^2 &=& \frac{1}{1-\mu^2\rho^2\frac{L^2}{3}\cosh^2\left[\frac{3 t r_0}{2L^2}\right]},\nn\\
X_7^i &=&|X_7| = X_0 + \mu^2(X_0L^2)^2\label{Xs},
\eea
in four dimensions. Finally then, the four dimensional metric is
\bea\label{full_MD}
  ds_4^2 &=&-dt^2\left(\frac{3r_0^2}{4L^2}\rho^2+
  \frac{\mu^2\rho^4r_0^2}{4}\cosh^2\left[\frac{3 t r_0}{2L^2}\right]\sinh^2\left[\frac{3 t r_0}{2L^2}\right]  
  \right) \cr
  &&+ d\rho^2\frac{L^2}{3}\left(1-\mu^2\frac{L^2}{3}\rho^2\cosh^4\left[\frac{3 t r_0}{2L^2}\right]\right)
  \ret &&-d\rho \, dt \, \mu^2 r_0 \frac{L^2}{3}\rho^3 \cosh^3\left[\frac{3 t r_0}{2L^2}\right]
  \sinh\left[\frac{3 t r_0}{2L^2}\right] + \frac{L^2}{r_0^2} dx_1^2 + dx_{11}^2.
\eea
Some points about this metric deserve further clarification. Obviously, as we claimed earlier, (\ref{mmetric}) is not of the reduced type (\ref{redansatz}). However, because 
the two match up to quadratic order in the internal $y$ coordinates\footnote{Note that the equations of motion only involve second derivatives, 
hence in the neighbourhood of the North Pole we only need to have the solution up to quadratic order in $y$ in order to satisfy them.}, 
there is a solution that looks like the reduction ansatz 
(\ref{redansatz}) and whose four-dimensional metric is a small perturbation of (\ref{MD}). 
We also note that we only need to define the (arbitrary) initial solution on a {\it Cauchy surface}, and so we needn't worry about satisfying the {\it exact} equations of motion for the initial solution that we define here. 

Returning to the original coordinate $u$ with $\rho=2(1/u-1)^{1/2}$, we have
\bea\label{mass_deformed_metric}
ds^2&=& ds_{AdS_4\,BH}^2 + \mu^2 ds_{\rm MD}^2,
\eea
where
\bea\label{MD}
ds_{\rm MD}^2 &=& -dt^2 r_0^2\frac{(1-u)^2}{u^2}\cosh^2\left(\frac{3 t r_0}{2L^2}\right)
\sinh^2\left(\frac{3 t r_0}{2L^2}\right) 
- du^2\frac{4L^4}{9u^4}\cosh^4\left(\frac{3 t r_0}{2L^2}\right)\ret
&&-du \, dt \frac{8 L^2r_0(1-u)}{3u^3} \cosh^3\left(\frac{3 t r_0}{2L^2}\right)\sinh
\left(\frac{3 t r_0}{2L^2}\right). 
\eea

This metric \eqref{mass_deformed_metric}
is our first order solution that we now employ as the {\it initial data} to the  Einstein's equations to 
obtain the back-reacted metric and Weyl tensor at the horizon subject, of course, to the caveat that the solution corresponds to the mass deformation of the horizon only near $t=0$.

Note that we now have a metric with scalars turned on near the horizon. Thus, our Einstein-Maxwell action with the Weyl coupling term needs to be further 
modified in order to account for the backreaction of the scalar fields. As argued in the introduction, we do not have the luxury of having an 
exact black hole solution for the gravity dual of the mass-deformed ABJM theory. What we do have, however, is the leading order mass deformation at the horizon, sourced by the scalars $T_{AB}$, which is sufficient to compute the membrane DC conductivity. 

In order to find the correct coupling of the scalars to gravity and the Maxwell fields, we focus on the bosonic part the of the four-dimensional gauged supergravity action. We are therefore led to consider the action
\bea\label{EMWGS}
I &=& \int d^4x \sqrt{-g} \left[R- \frac{1}{2} m^2 \left({\rm Tr}(T)^2-2 {\rm Tr}(T^2)\right) - 
\frac{1}{4} {\rm Tr} \left(\partial_{\mu}T^{-1}\partial^{\mu}T\right) \right. \nn\\
&&\left. +\frac{1}{g_4^2}\left(-\frac{1}{4}T_{AB}F_{\mu\nu}^A F^{B\mu\nu} + 
\gamma L^2 C_{\mu\nu\rho\sigma}F^{A\mu\nu}F^{A\rho\sigma}\right)\right] 
\eea
with $T_{AB}\equiv L^2 X_A \delta_{AB}$, with $A=0,\cdots, 7$. The $L^2$ factor is included in the definition of the $T_{AB}$ fields for dimensional reasons. We close this subsection by writing down the Einstein's equations that follow from the action \eqref{EMWGS} which we will use 
to compute the membrane conductivity. Recalling that we are considering the situation where the gauge fields are taken as probes, that is, that the 
putative black hole background is not modified by the gauge perturbation, the Einstein's equations that follow from \eqref{EMWGS} are:
\bea\label{eom}
R_{\mu\nu}=\mathcal{T}_{\mu\nu}-\frac{1}{2}g_{\mu\nu} \mathcal{T},
\eea
where the energy momentum tensor $\mathcal{T}_{\mu\nu}$ given by
\bea
\cal{T}_{\mu\nu} &=& \frac{1}{8} {\rm Tr} (\partial_{\mu} T^{-1} \partial_{\nu} T+
\partial_{\nu} T^{-1} \partial_{\mu} T) - g_{\mu\nu} U \, , 
\eea
with  $\mathcal{T}=\mathcal{T}^{\mu}{}_{\mu}$ , and
\bea
U \equiv 
\frac{1}{4}m^2\left({\rm Tr}(T)^2-2{\rm Tr}(T^2)\right)+\frac{1}{2}{\rm Tr} (\partial_{\rho} T^{-1} \partial^{\rho} T).
\eea
As a final remark we should also give the relation between the parameter $m^2$ and the $AdS$ scale $L.$ 
The most straightforward way to understand this relation is to compare the Ricci scalars of the 
pure $AdS_4$ black hole and the mass-deformed one.  In the pure case, $R=4\Lambda=-12/L^2$, whereas 
with the mass deformation turned on, we have $R=-\mathcal{T}$.
In the $\mu\to 0$ limit, from \eqref{scalars} we have that $T_{AB} \to \delta_{AB}$, giving 
$R \to 48m^2$. Since this has to match with $-12/L^2$, we obtain\footnote{Note that this squared-mass is 
well above the Breitenlohner-Freedman bound.}
\bea
m^2=-\frac{1}{4L^2} \,.
\eea 

\subsection{Integrating the Einstein's equations to get higher order deformations}

Now that we have obtained a first order solution, we need to use the Einstein's equations to find the required second- and higher-order solution near the 
horizon. Indeed, in the action (\ref{mxaction}), we have not just the near-horizon metric, but the Weyl tensor as well, containing second order derivatives
of the metric. To specify a solution of the Einstein's equations, which are second order differential equations, we need to specify the metric and its first derivative
(in a non-tangential direction) on a codimension-one Cauchy surface. Therefore, if the coordinate away from the surface, i.e. the foliation direction, 
is called $v$, and the surface is at $v=0$, we need to specify the two sets of functions $g_{\mu\nu}(\vec{x},v=0)$ and $\d_v g_{\mu\nu}(\vec{x},v=0)$, 
with $\vec{x}$ coordinates tangent to the Cauchy surface. Then the Einstein's equations determine the full solution by integration, and in particular, 
we can obtain at least some of the second order derivatives algebraically (without the need to integrate them).
We are accustomed with the situation when $v$ is actually time, and we use a time foliation such that the Cauchy surface is at $t=0$, but in our case 
we are actually interested in the situation when $v$ is the radial direction away from the horizon, $\rho$, and the Cauchy surface is at the horizon, 
$\rho=0$.

This situation is a bit subtle, since components of the metric are actually divergent or zero at the horizon, hence to check that we are applying it 
correctly, we first test it on an example when we actually know the full solution, the AdS black hole of (\ref{adsbh}). We will keep only the 
zeroth and first order solution, and deduce the second order, and in particular the Weyl tensor, from the Einstein's equations.

As mentioned before, we already know the full metric \eqref{adsbh}, but we will take it in the near horizon limit ($u\rightarrow 1$) and its first derivative as the only information given. Note however, that $g_{tt}$ is zero at the horizon, whereas $g_{rr}$ is infinite.
So we have non-zero components of the metric that behave as
\bea
g_{\a\b}=\frac{g^{(-1)}_{\a\b}}{(u-1)} + g^{(0)}_{\a\b} + g^{(1)}_{\a\b}(u-1) + {\cal O}(u-1)^2,\label{metnh}
\eea
The first derivatives with respect to the coordinate $u$ will be denoted with a prime ($g'_{\a\b}$). In this case therefore we see that giving the metric and
its first derivative (as functions of the remaining coordinates) on a Cauchy surfaces needs to be replaced by giving the first two coefficients. For $g_{rr}$ 
this means $g^{(-1)}$ and $g^{(0)}$, whereas for the rest it is $g^{(0)}$ and $g^{(1)}$. We should also comment on a more general case (even though we are not aware of such examples for the Einstein-Hilbert action): if it happens that some component starts at Laurent expansion order $g^{(-p)}$, we would give $g^{(-p)}$
and $g^{(-p+1)}$, whereas if it starts at $g^{(p)}$, we would need to specify at least $g^{(p)}$ ($g^{(p-1)}=0$ could be a valid specification, depending on the
case at hand). In the case of the AdS black hole test solution, we will take these 
first two coefficients in the components of $g_{\a\b}$  as known, but we will use their explicit expressions just at the end of the calculations.

For the Einstein's equations, and later for the Weyl tensor, the primary ingredient is the Riemann tensor. It is practical to write the Riemann tensor with the second derivatives of the metric explicitly identified, that will make solving of the algebraic equations simpler. 
The expression for the Riemann tensor that will be used is
\bea
R_{\a\b\g\delta}=\frac{1}{2}(\pd_{\a\delta}g_{\b\g} - \pd_{\a\g}g_{\b\delta} + \pd_{\b\g}g_{\a\delta} - \pd_{\b\delta}g_{\a\g})+\G^{\e}_{\a\delta}\G_{\b\e\g} - \G^{\e}_{\a\g}\G_{\b\e\delta},\label{rtens}
\eea
where, by definition, the Christoffel symbols contain first derivatives only.

Let us recall the Einstein equations with a cosmological constant and extract one piece of information before the full calculation. They are
\bea
R_{\a\b} - \frac{1}{2}g_{\a\b}R + g_{\a\b}\L = 0. \label{eeqs} 
\eea
We can take the trace in the previous equation and obtain the curvature scalar without major effort. In 
four dimensions it is $R = 4\L$. For the full metric the Ricci scalar is $-\frac{12}{L^2}$, and combining these last two 
expressions yields $\L=-\frac{3}{L^2}$.

The Ricci tensor is now
\bea
R_{\a\b}=\L g_{\a\b}, \label{algeqs}
\eea
which contains inside the second derivatives of the metric (so from here we will have our algebraic equations) and can also be used to calculate the Schouten tensor. In four dimensions, the Schouten tensor has the following expression
\bea
S_{\a\b} = \frac{1}{2}\left(R_{\a\b} - \frac{R}{6}g_{\a\b}\right).
\eea
Using the Einstein's equations here will lead to $S_{\a\b}=\frac{\L}{6}g_{\a\b}$. The Schouten tensor is useful in this 
case to simplify the calculation of the Weyl tensor (it also contains all the matter information, as we will see later), which can be written as
\bea
{C_{\a\b}}^{\g\delta}={R_{\a\b}}^{\g\delta} - 4S^{[\g}_{[\a}\delta^{\delta]}_{\b]}. 
\eea

Let us calculate the Weyl tensor component ${C_{uy}}^{uy}$ that was used to obtain the conductivity in the massless case. 
Here we have
\bea
{C_{uy}}^{uy}={R_{uy}}^{uy} - \frac{\L}{3}. 
\eea
The Einstein's equations will play its role here as the relation to find ${R_{uy}}^{uy}$ algebraically. We will have to look at the equation for the component $R_{yy}$ of the Ricci tensor,
\bea
R_{yy} = g^{tt}R_{tyty} + g^{uu}R_{uyuy} + g^{xx}R_{xyxy}.
\eea  
We go then to equation \eqref{algeqs} for $R_{yy}$, and rising the $y$ index we obtain
\bea
{R_{uy}}^{uy}=\L - g^{yy}(g^{tt}R_{tyty} + g^{xx}R_{xyxy}).
\eea
Using \eqref{rtens} and the assumption that we have a static and spherically symmetric black hole solution  -- which is 
in fact true for the full solution \eqref{adsbh} --, we see that $R_{uy}{}^{uy}$, and therefore $C_{uy}{}^{uy}$, are 
determined solely in terms of the metric and its first derivatives.
After adding everything up and replacing the respective values of the metric and its first derivatives, we have the 
following expression for the sought Weyl tensor component
\bea
{C_{uy}}^{uy}=- \frac{1}{2L^2} + \mathcal{O}(1-u)\label{C_massless},  
\eea
in the near horizon limit. This is in complete agreement with the result one obtains from the full solution (\ref{adsbh}) 
given by
\bea
C_{uy}{}^{uy} = -\frac{u^3}{2L^2}.
\eea

\subsection{The higher order deformation at the horizon}

We now move to apply the procedure defined and tested in the last subsection to our four dimensional mass-deformed metric near the horizon of the 
black hole.

This case will present more ingredients than the previous one, but the method to obtain the second derivatives is applied in the same way. The massive deformation for the $AdS_4$ black hole enters in the metric like
\bea
ds^2=ds^2_{AdS_4-BH} + \m^2ds^2_{MD}
\eea
where $ds^2_{AdS_4-BH}$ is the near horizon expansion of the form \eqref{metnh} of \eqref{adsbh} and the massive deformation is given in 
\eqref{MD}.

Now the desired component of the Weyl tensor is $C^{uyuy}$ (see eq.\eqref{sigma_JvsF}) which, written 
in terms of the Riemann and Schouten tensors components is 
\bea
C^{uyuy} &=& (g^{yy})^2\left((g^{uu})^2R_{uyuy} + 2g^{uu}g^{tu}R_{tyuy} + (g^{tu})^2R_{tyty} \right) \cr 
&&- g^{yy}\left(g^{uu}({S^u}_u + {S^y}_y) + g^{tu}{S^u}_t \right). 
\label{weylmd}
\eea

Like in the previous case, assuming also that we have spherical symmetry in the full solution 
(i.e. the metric does not depend on $y$, nor $x$) only the $R_{uyuy}$ component involves the 
second derivatives $\partial^2_u g_{yy}$; the other components $R_{tyuy}$ and $R_{tyty}$ 
only contain the metric and its first derivatives. 

But there is a caveat: we need to assume that $g_{yy}$ is $t$-independent in the full solution, 
not only in the Cauchy solution, if not, we could have unknown second time derivatives of the metric to calculate as well. We believe however that these are reasonable assumptions, though at 
this moment we cannot rigorously prove them.

The Einstein's equations \eqref{eom} now involve an energy-momentum tensor. From those 
equations we will now obtain the Schouten tensor, that will capture just matter content
\bea
S_{\a\b}=\frac{1}{2}\left({\cal T}_{\a\b} - \frac{1}{3}{\cal T}g_{\a\b}\right).
\eea

Let us write again the Ricci tensor component that we need for this case,
\bea
R_{yy} = g^{tt}R_{tyty} + g^{uu}R_{uyuy} + 2g^{tu}R_{tyuy} + g^{xx}R_{xyxy}.
\eea

From here we see that the combination $g^{uu}R_{uyuy} + 
2g^{tu}R_{tyuy}$, that also appears in the Weyl tensor \eqref{weylmd}, can be written in terms of 
$R_{tyty}$ and $R_{xyxy}$ which do not involve any second derivatives of the metric; so only these 
two components need to be computed. Thus, we have managed to write $C^{uyuy}$ completely 
in terms of the metric and its first derivatives as desired.

Making all the replacements in \eqref{weylmd} the component for the mass deformed Weyl tensor in the near horizon limit reads
\bea
C^{uyuy} =  \left(\frac{3}{2r_0^2L^2} + \mu^2 \frac{21}{2r_0^2}\right)(u-1).
\eea
As a consistency check we see that, after using \eqref{adsbh} also, the $\mu\to0$ limit above 
gives the correct answer for the massless case \eqref{C_massless}.

\section{Conductivity of the mass-deformed system}
\label{Sec:Conductivity}

We finally return to the calculation of the conductivity. There are two (equivalent) ways to calculate it using the membrane paradigm, 
as we saw in section \ref{sec:membrane}: one, using a version of Kubo's formula at the horizon, and the other using formula (\ref{conductivity}), obtained by adding a surface term involving the current.

There is one more observation to be made: the conductivity is usually defined with respect to a $U(1)$ (abelian) field, whereas we have an $SO(8)$ (nonabelian) field, 
so the conductivity will depend on how we embed the abelian field inside $SO(8)$. In principle there are many ways to do this but we saw that, due to our approximation
of being near the North Pole of the $S^7$ of the compactification that generates $SO(8)$, the diagonal gauge field associated with $X^0$  ($A^0$, for index $A=0$) is 
special, and different than the diagonal gauge fields associated with $X^i$ ($A^i$, for $A=i$). At the horizon, the $X^i$ scalars ``feel'' the massive 
deformation while the $X^0$ field does not (see equation (\ref{nh_Xs})).

From  section \ref{sec:membrane}, it follows that we need only compute 
the effective coupling $g^2_{ry,4}$ at the horizon (see eq. \ref{effmaxwell}) with the mass 
deformation now turned on. In this case,
\bea
g^{uu} &=& \frac{3(1-u)}{L^2}+\mathcal{O}(1-u)^2\, , \quad 
g^{yy} = \frac{L^2}{r_0^2} +\mathcal{O}(1-u),
\label{nh_metric}\nonumber\\
C^{uyuy} &=& \left(\frac{3}{2r_0^2L^2}+\mu^2\frac{21}{2r_0^2}
\right)(u-1)+\mathcal{O}(1-u)^2\label{nh_Weyl},\\
X_0&=& \frac{1}{L^2} + \mathcal{O}(1-u)\, , \quad  
X_i=\frac{1}{L^2}+\mu^2 + \mathcal{O}(1-u)\label{nh_Xs},\nonumber
\eea
near the horizon. Concentrating on the gauge field part of the action (\ref{EMWGS}), for $T_{AB}=L^2X_A\delta_{AB}$, the terms involving $F_{ry}$ are 
\be
-\frac{1}{4}\int d^4x \sqrt{-g}F^A_{ry}F^{A,ry}\Big[L^2X_A-8\gamma {C^{ry}}_{ry}\Big].
\ee
Therefore using \eqref{conductivity} and the 
near horizon behavior listed in equations \eqref{nh_metric} through \eqref{nh_Xs}, we find that
\bea
\sigma^{(0)}&=&\frac{1}{g^2_4}\left(1+4\gamma + 28 \gamma \mu^2 L^2\right)\\
\sigma^{(i)}&=&\frac{1}{g^2_4}\left(1+\mu^2L^2+4\gamma 
\left(1+ 7 \mu^2 L^2\right)\right),
\eea
where the $(0)$ and $(i)$ superscripts on the left hand side denote the conductivities associated 
with the $A^0_{\mu}$ and $A^i_{\mu}$ gauge fields respectively.\footnote{Note that $\mu$ corresponds exactly to the mass parameter in the gauge theory, 
as already noted. Then in the gauge theory, the corrections to the dimensionless conductivity (defined at zero frequency and momentum) must also come in a 
dimensionless combination of $\mu$ and a quantum scale in the nonconformal massive ABJM theory, corresponding to $L$, just like in the "hard-wall" model for
QCD of Polchinski and Strassler, and in other theories with a mass gap that are conformal in the UV, $L$ corresponds to $\Lambda_{\rm QCD}^{-1}$.  
Also, $g_4$ was related to gauge theory parameters in (\ref{g4}) and $\gamma$ is a semi-phenomenological parameter from the point of view of gauge theory.}
One potential problem that we alluded to earlier was that now we also have a non-vanishing off-diagonal term in the metric $g_{rt}$. However, as we show in Appendix B, its effects cancel each other out at the horizon and do not contribute in the case of a scalar field. In Appendix C, we go one step further and show also that in the case of the gauge field, the above calculation for the conductivity goes through without modification due to a similar type 
of cancellations that occur at the horizon. Notice also that since $X^0$ does not ``see'' the mass deformation at the horizon, 
$\sigma^{(0)}$ is blind to the effect of $\mu^2$ as we turn off the Weyl coupling $\gamma$.
 
Finally, we can use the Kubo formula for the boundary term at the horizon, just as in the $\mu=0$ case. Here, we have $X^{uyty}$ nonzero in the 
background, so the action for the $A_y$ field is now
\be
  S_{yy} = -\frac{1}{2g_4^2}\int d^3x \sqrt{-g} \left(X^{uyuy}A_y \partial_{u}A_y + 
  X^{uyty}A_y \partial_t A_y\right)\bigg|_{u\to 1}.
\ee
As before, the associated Green's function  
\be
  G_{yy}^{(A)} = -\frac{3r_0}{L^2g_4^2}\left(1+4\gamma+h_A(\mu^2)\right)(1-u)
  \frac{\partial_u A_y^A(\omega)}{A_y^A(\omega)}\bigg|_{u\to 1},
\ee
where the index $A$ is used to describe the gauge field whose conductivity we are calculating, leading to
\bea
h_0(\mu^2)&=&28\gamma \mu^2 L^2 \cr
h_k(\mu^2)&=&\mu^2L^2+28\gamma \mu^2 L^2\label{hA}.
\eea
The equation of motion for $A_y$ near the horizon is then
\be
  \frac{\partial_u A_y(\omega)}{A_y(\omega)}\bigg|_{u\to 1} =  
  i\omega\frac{L^2}{3r_0}\frac{1}{1-u}+i\omega {\cal O}((1-u)^0).
\ee 
Consequently then,
\bea
  \sigma^{(A)} = \frac{1}{g^2_4}\left(1+4\gamma+h_A(\mu^2)\right).
\eea
More details of this computation may be found in Appendix D.

\section{Conclusions}

To summarize then, in this article we have analyzed the effect of a massive deformation of the ABJM model on the conductivity calculated from the gravity dual. However, unlike 
the computation in \cite{Myers:2010pk}, since we don't have the full $A_y$ defined from the horizon to the boundary, nor even an explicit formula for the gravity dual, we need to use membrane paradigm-type calculations. In particular we have developed a novel calculation based on an extension of the Kubo formula, only using a 
boundary term at the horizon, together with a corresponding extension of the membrane paradigm method presented in \cite{Iqbal:2008by}. 

Crucial to a computation of the conductivity is knowledge of the mass-deformed black hole solution. However, given the complications of the background dual to the mass-deformed ABJM model, such an exact solution remains unknown. Nevertheless, we were able to compute the conductivity knowing only the effect of the mass deformation on the near horizon region of the $AdS_{4}$ black hole. In order to determine the mass deformation of the solution near the horizon, we employed a two-step process: First, we input the zeroth and first 
order solutions as Cauchy data in the Einstein equations, and to find the correct ones that correspond to the mass deformation, superposed a pp-wave on the (approximately flat) horizon and T-dualized. Then, in order to find the higher order solution near the horizon, we developed a method of using Einstein's equations and the Cauchy data. This procedure was benchmarked against the known AdS black hole solution with excellent agreement. Finally, we obtained the same value for the mass-deformed DC conductivity using both membrane paradigm calculations, with both
exhibiting an increase as a function of $\mu^2$. 

Of course, given the motivation for this work, we would have wanted to obtain $\sigma(\omega)$ at nonzero $\mu$, with the hope that the mass deformation could replace the 
deformation by the Weyl tensor coupling $\gamma$ at least in some regime (which should include $\mu\ll T$ so that we still have an approximate 
conformal field theory at finite temperature). But, due to the technical complications of the background geometry illustrated above, we could only calculate the DC conductivity, $\sigma(0)$. In that respect, all we are able to say is that the mass deformation gives a positive contribution, like a positive $\gamma$, but are unable to speculate any further on what happens at nonzero $\omega$. It goes without saying that it would be of enormous interest to extend the calculation to finite $\omega$ in the future.

\section{Acknowledgements}

The work of HN is supported in part by CNPq grant 301219/2010-9 and work of FR is supported by 
FAPESP grant 2012/05451-8. CLA would like to thank CAPES for full support. JM acknowledges support from the NRF of South Africa under the HCDE and IPRR programs. FR would like to thank  Ido Adam, Ilya Bakhmatov, 
Oscar Chacaltana, and especially Alfonso Ballon-Bayona for very helpful discussions.

\appendix
\section{T-duality from IIB to IIA}\label{App:Tduality}

In this appendix we will give a few details of the T-dualization of the mass-deformed metric. Expanding out the differentials in \eqref{IIBsolution}, the nonzero components of the metric are:	
\bea
g_{\tau\tau}&=&-\bar{\rho}^2 \left(1+\mu ^2 \cosh^2\tau\left(\bar{\rho} ^2 \cosh^2\tau
+y_7^2\right)\right)\\
g_{\bar{x}_1\bar{x}_1}&=&1-\mu ^2 \left(\bar{\rho} ^2 \cosh^2\tau+y_7^2\right)\\
g_{\bar{\rho}\bar{\rho}}&=& 1-\mu ^2 \sinh^2\tau \left(\bar{\rho} ^2 \cosh^2\tau+y_7^2\right)\\
g_{\bar{\rho}\tau}&=&-\mu ^2 \bar{\rho}  \cosh\tau\sinh\tau \left(\bar{\rho} ^2 \cosh^2\tau+y_7^2\right)\\
g_{\bar{x}_1\tau}&=&-\mu^2 \bar{\rho} \cosh\tau (\bar{\rho}^2 \cosh^2\tau +y_7^2)\\
g_{\bar{\rho}\bar{x}_1}&=&-\mu ^2 \sinh\tau \left(\bar{\rho} ^2 \cosh^2\tau+y_7^2\right)
\eea	
Now we will T-dualize on the $\bar{x}_1$ direction to IIA, application of the Buscher rules yields
\bea
\tilde{g}_{\bar{x}_1\bar{x}_1}&=& 1/g_{\bar{x}_1\bar{x}_1}=\frac{1}{1-\mu ^2 \left(\bar{\rho} ^2 \cosh^2\tau+y_7^2\right)}\\
\tilde{g}_{\tau\tau}&=&g_{\tau\tau}-(g_{\tau \bar{x}_1})^2/g_{\bar{x}_1\bar{x}_1} 
= -\frac{\bar{\rho}^2\left(1+\mu^2\sinh^2\tau\left(\bar{\rho} ^2 \cosh^2\tau+y_7^2\right)\right)}
{1-\mu ^2 \left(\bar{\rho} ^2 \cosh^2\tau+y_7^2\right)}\\
\tilde{g}_{\bar{\rho}\bar{\rho}}&=&g_{\bar{\rho}\bar{\rho}}-
(g_{\bar{\rho}\bar{x}_1})^2/g_{\bar{x}_1\bar{x}_1} 
=\frac{1-\mu^2 \cosh^2\tau \left(\bar{\rho} ^2 \cosh^2\tau+y_7^2\right)}
{1-\mu ^2 \left(\bar{\rho} ^2 \cosh^2\tau+y_7^2\right)}\\
\tilde{g}_{\bar{\rho}\tau}&=&g_{\bar{\rho}\tau}-
(g_{\bar{\rho}\bar{x}_1}g_{\tau\bar{x}_1})/g_{\bar{x}_1\bar{x}_1} 
=-\frac{\mu^2 \bar{\rho}\cosh\tau\sinh\tau \left(\bar{\rho} ^2 \cosh^2\tau+y_7^2\right)}
{1-\mu ^2 \left(\bar{\rho} ^2 \cosh^2\tau+y_7^2\right)}\\
\tilde{\phi}&=& -\frac{1}{2}\log|g_{\bar{x}_1\bar{x}_1}|=-\frac{1}{2}\log\left(1-\mu ^2 \left(\bar{\rho} ^2 \cosh^2\tau+y_7^2\right)\right)\\
\tilde{B}_{\tau\bar{x}_1}&=&g_{\tau\bar{x}_1}/g_{\bar{x}_1\bar{x}_1} = 
\frac{-\mu^2 \bar{\rho} \cosh\tau (\bar{\rho}^2 \cosh^2\tau +y_7^2)}
{1-\mu ^2 \left(\bar{\rho} ^2 \cosh^2\tau+y_7^2\right)}\\
\tilde{B}_{\bar{\rho}\bar{x}_1}&=&g_{\bar{\rho}\bar{x}_1}/g_{\bar{x}_1\bar{x}_1} = 
\frac{-\mu ^2 \sinh\tau \left(\bar{\rho} ^2 \cosh^2\tau+y_7^2\right)}
{1-\mu ^2 \left(\bar{\rho} ^2 \cosh^2\tau+y_7^2\right)}.
\eea
Then, the mass-deformed metric on type IIA has the following form:
\bea
ds^2_{IIA}&=&\frac{1}{1-\mu ^2 \left(\bar{\rho} ^2 \cosh^2\tau+y_7^2\right)}
\Big(-\bar{\rho}^2d\tau^2 + d\bar{\rho}^2 + d\bar{x}_1^2\cr
&& - \mu ^2 \left(\bar{\rho} ^2 \cosh^2\tau+y_7^2\right)(d(\bar{\rho}\cosh\tau))^2\Big) 
+d\vec{y}_7^2.
\eea

\section{Membrane paradigm in the presence of $g_{rt}$}

In the metric we obtain, we have an off-diagonal component $g_{rt}$, so in this Appendix we study its effect on the formulas for the membrane paradigm.

\subsection{Infalling condition}

We would like to find the infalling boundary condition for a metric of the form 
\be\label{metricul}
ds^2=-g_{tt}(r,t)dt^2 +g_{rr}(r,t) dr^2 +2g_{rt}(r,t)dr\,dt+ g_{yy}(r)(dx^2+dy^2). 
\ee 
Define first the tortoise coordinate $r^*$ such that $ds^2$ can be written as
\be
ds^2= dv du + g_{yy}(r)(dx^2+dy^2)
\ee 
where $dv= dr^*+dt$, $du=dr^*-dt$. To find $dr^*$ in terms of $dr$ we write
\bea
dv&=& dt+dr^*\equiv dt+a\,dr\label{v}\\
du&=&dt-dr^*\equiv b\,dt + d\,dr
\eea
which gives a quadratic equation for $a$, namely, $a^2 g_{tt}+2g_{rt}a-g_{rr}=0$. Solving for $a$ yields  
\bea\label{tortoise}
dr^*= a\,dr=\left(-\frac{g_{rt}}{g_{tt}}\pm \sqrt{\frac{g_{rt}^2}{g_{tt}^2}+\frac{g_{rr}}{g_{tt}}}\right)dr. 
\eea
Consider now a scalar field $\phi$ near the horizon of a black hole described by the metric \eqref{metricul}. The infalling 
condition at the horizon means that $\phi$ can only depend on $r$ and $t$ through the non-singular combination given 
by the Eddington-Finkelstein coordinate $v$, defined in \eqref{v}. Thus, for fixed $v$,  we have $dt=-a dr$, and
\be
d\phi(v) = \partial_r \phi(r,t)dr+\partial_t \phi(r,t)dt =
\left(\partial_r \phi(r,t)-a\partial_t \phi(r,t)\right)dr = 0
\ee
yielding the relation 
\bea\label{EF}
\partial_r \phi(r,t)=a\, \partial_t \phi(r,t)
\eea 
but now with $a$ as defined in \eqref{tortoise}. Note that in order 
to recover the usual infalling condition $\partial_r\phi =\sqrt{g_{rr}/g_{tt}}\,\partial_t\phi$ when 
$g_{rt}=0$, we need to take the $+$ solution in \eqref{tortoise}.

\subsection{Scalar field calculation}

To see the effect of the off-diagonal metric on a membrane paradigm-type calculation, we look at the example of a scalar field.

Consider a massless bulk scalar field with action
\bea
S=-\frac{1}{2}\int_{r>r_0} d^{d+1}x\sqrt{-g}\frac{1}{g^2_{d+1}(r,t)}\partial_{\mu}\phi\partial^{\mu}\phi
\eea
where the horizon is at $r=r_0$, and we now have allowed for an $r$ and $t$ dependent scalar coupling $g^2_{d+1}(r,t)$. This 
is to account for the fact that our background fields can now depend on $t$ as well (see for example 
equations \eqref{Xs} and \eqref{MD}). Following \cite{Parikh:1997ma}, we need to add a boundary term at 
the horizon to the one that arises from variations of 
this action.  This term is given by
\bea
S_{\rm surf} = \int_{\Sigma}d^dx \sqrt{-\gamma} \left(\frac{\Pi(r_0,x)}{\sqrt{-\gamma}}\right)\phi(r_0,x)
\eea
where $\gamma_{\mu\nu}$ is the metric induced at the stretched horizon $\Sigma$, and $\Pi$ is the momentum conjugate to $\phi$ with respect to a foliation in the $r$-direction, i.e.: 
\bea
\Pi \equiv \frac{\partial \mathcal{L}}{\partial(\partial_r \phi)} = -\frac{\sqrt{-g}}{g^2_{d+1}(r,t)} g^{r\mu}\partial_{\mu}\phi.
\eea
The ``membrane $\phi$-charge'' is therefore
\bea
\Pi_{\rm mb} \equiv \frac{\Pi(r_0,x)}{\sqrt{-\gamma}} = - \frac{1}{q(r,t)}
\left(g_{rr}+\frac{g_{rt}^2}{g_{tt}}\right)^{1/2}\left(g^{rr}\partial_{r}\phi+g^{rt}\partial_{t}\phi\right)
\eea
where in the last equality we have used the form of the metric \eqref{metricul}. Now we use the infalling 
condition \eqref{EF}, obtaining
\bea
\Pi_{\rm mb} &=& - \frac{1}{g^2_{d+1}(r,t)}
\sqrt{\frac{\Delta}{g_{tt}}}
\left(-\frac{g^{rr}g_{rt}}{g_{tt}}+ \frac{g^{rr}}{g_{tt}}\sqrt{\Delta}+g^{rt}\right)\partial_{t}\phi\cr
&=&- \frac{1}{g^2_{d+1}(r,t)} \frac{1}{\sqrt{g_{tt}}}\partial_t \phi
\label{Pimb}
\eea
where $\Delta\equiv g_{rr}g_{tt}+g_{rt}^2$. 
If go to the frame of an observer hovering just outside the horizon with proper time $\tau$, then
\be
\Pi_{\rm mb}= - \frac{1}{g^2_{d+1}(r,t)} \partial_{\tau} \phi.
\ee
Therefore, the effect of the off-diagonal metric component $g_{rt}$ gets completely cancelled out and the 
membrane response $\Pi_{\rm mb}$ is the same as in the case of a diagonal metric (see e.g. \cite{Iqbal:2008by}).

\section{Conductivity using $J^i$ vs. $F_{ti}$ relation in the presence of $g_{rt}$}

Consider now a generalization of the action in (\ref{xffaction}), namely
\bea
S &=&\int d^4x \sqrt{-g}\left[ -\frac{1}{8g^2_4}X_{AB}^{\mu\nu\rho\sigma}F^A_{\mu\nu}F^B_{\rho\sigma}\right]
\eea
with
\bea
X_{AB\, \mu\nu}{}^{\rho\sigma} &\equiv& 2\delta_{[\mu\nu]}{}^{[\rho\sigma]} T_{AB}-
8\gamma L^2 \delta_{AB}C_{\mu\nu}{}^{\rho\sigma}. 
\eea
Generalizing the result of section 3, we need to add the surface term
\bea
S_{\rm surf} = \int_{\Sigma}d^3x \sqrt{-\gamma}\left(\frac{j^{\mu}_A}{\sqrt{-\gamma}}\right)A_{\mu}^A
\eea
with 
\bea
j^{\mu}_A \equiv \frac{\partial \mathcal{L}}{\partial(\partial_r A_{\mu}^A)} = -\frac{1}{2g^2_4}\sqrt{-g} X^{r\mu\nu\rho}_{AB}F_{\nu\rho}^B.
\eea
Thus, the membrane current in this case is 
\be
J^{\mu}_{\rm mb} \equiv \left(\frac{j_A^{\mu}}{\sqrt{-\gamma}}\right) 
=-\frac{1}{2g^2_4}
\frac{1}{\sqrt{g^{rr}}}X^{r\mu\nu\rho}_{AB}F_{\nu\rho}^B\Big|_{r=r_0}.
\ee
Using the ansatz \eqref{metricul}, the only non-zero components of the $X$ 
tensor that enter in the right hand side above for $J^i_{\rm mb}$ are $X^{riri}$ and $X^{riti}$, yielding
\bea
J^{i}_{\rm mb}
&=&-\frac{1}{g^2_4}
\frac{1}{\sqrt{g^{rr}}}\left(X^{riri}_{AB}F_{ri}^B+X^{riti}_{AB}F_{ti}^B\right)\Big|_{r=r_0}
\eea
(no sum over $i$). Now, for the $A_i$ component of the gauge fields (we omit the scalar indices 
$A,B$  for now), the infalling condition \eqref{EF} at the horizon is 
\bea\label{EFA}
\partial_r A_i = a \, \partial_t A_i \,\, , \quad r\to r_0
\eea
which, using the $A_r=0$ gauge and the condition that $A_t=0$ vanishes at the horizon\footnote{This is required 
in order to have a nonsingular gauge connection (see, for example, \cite{Hartnoll:2011fn})}, yields 
\bea
F_{ri} = a\, F_{ti} \quad {\rm as} \,\, r \to r_0.
\eea
This implies that $J^i_{\rm mb}$ is proportional to $F_{ti}$ at the horizon as expected, i.e.
\bea
J^{i}_{\rm mb}
&=&-\frac{1}{g^2_4}
\frac{1}{\sqrt{g^{rr}}}\left(a X^{riri}+X^{riti}\right)F_{ti}\Big|_{r=r_0}.
\eea
The factor in parenthesis after some algebra becomes
\bea
a X^{riri}+X^{riti} = \Delta^{-1/2} g^{ii} - 8\gamma L^2\left(\frac{\Delta^{1/2}}{g_{tt}}C^{riri}-\frac{g_{rt}}{g_{tt}}C^{riri}+C^{riti}\right)\label{aXX}.
\eea
For our ansatz \eqref{metricul}, we have
\bea
\frac{C^{riti}}{C^{riri}}= \frac{g_{rt}}{g_{tt}} \, ,
\eea 
which implies that the last two terms inside the parentheses in \eqref{aXX} cancel each other 
out, giving
\bea
a X^{riri}+X^{riti} &=& \Delta^{-1/2} g^{ii} - 8\gamma L^2 \frac{\Delta^{1/2}}
{g_{tt}}C^{riri}.
\eea
After some more algebra, we arrive at 
\bea
J^i_{\rm mb} &=& -\frac{1}{g^2_4}\frac{\left(g^{rr}g^{ii}
-8\gamma L^2 C^{riri}\right)}{\sqrt{g_{tt}}g^{rr}}g_{ii}F_t{}^i
\eea
where we also used the fact that $g_{ij}=g_{ii}\delta_{ij}$ to write $F_{ti}=g_{ii}F_t{}^i$. 

In an orthonormal frame of a physical observer just outside the horizon, with proper time $\tau$, we 
have that $F_t{}^i=-\sqrt{g_{tt}}F^{\tau i}=
-\sqrt{g_{tt}}\hat{E}^i$, where $\hat{E}^i$ is the electric field measured by such observer. 

Reinserting the scalar indices $A,B$, and using $T_{AB}=L^2 X_A \delta_{AB}$, 
we arrive at a conductivity from the membrane paradigm given by
\bea
\sigma_{\rm mb}^A = \frac{L^2}{g_4^2}\left(X_A-\frac{8\gamma  g_{ii}C^{riri}}{g^{rr}}\right)\Big|_{r\to r_0}\label{sigma_JvsF}.
\eea
Using the formulas at the horizon (\ref{nh_metric}) to (\ref{nh_Xs}), we obtain
\bea
\sigma_{\rm mb}^{(0)} &=& \frac{1}{g_4^2}\left(1+4\gamma +28 \gamma\mu^2 L^2\right) 
\label{sigma0}\\
\sigma_{\rm mb}^{(k)} &=& \frac{1}{g_4^2}\left(1+4\gamma +\mu^2 L^2+ 
28 \gamma\mu^2 L^2\right)\label{sigmak}
\eea
which are the same we obtained using (\ref{conductivity}).

We can readily check that both of the formulas above give the correct result when the mass deformation is absent, namely \cite{Myers:2010pk}
\bea
\sigma_{\rm mb}^A \to \frac{1}{g_4^2}(1+4\gamma).
\eea

Recall that the relation we just obtained here corresponds to the conductivity defined on the horizon 
membrane. One way to relate to the conductivity at the boundary is to study the $r$ dependence of the current $j^i$ and electric field $F_{ti}$, 
by looking at the equation of motion and Bianchi identities involving $\pd_r j^i$ and $\pd_r F_{ti}$ 
along the lines of Appendix B in \cite{Iqbal:2008by}. 
Namely, in our case we have
\bea
\pd_r j^i &=& G X^{itri}\pd_t F_{ri}
+GX^{itti}\pd_t F_{ti}+ GX^{ijij}\pd_j F_{ij}\nonumber\\
&&+\pd_t(GX^{itri})F_{ri} +\pd_t(G X^{itti})F_{ti}\label{eomJ}\\
\pd_r F_{ti} &=& \pd_i F_{tr} + \pd_t F_{ri}\label{bianchi}
\eea
where $G\equiv \sqrt{-g}/g_4^2$ and $i\neq j$ (i.e., since the boundary is 2+1 dimensional, 
$i=y$ and $j=x$ or vice-versa). From here we immediately see that \eqref{bianchi} implies 
that $F_{ti}$ becomes $r$-independent in the $k_{\mu}\to 0$ limit, and that the first three terms in
\eqref{eomJ} will be suppressed in this limit. However, the last two terms in \eqref{eomJ} will 
remain finite since they do not involve space-time derivatives of the gauge fields, but only time derivatives of the metric. Since the near horizon metric 
\eqref{mass_deformed_metric}, \eqref{MD}, obtained by mass-deforming the flat horizon, does 
have an explicit time dependence -- at least for small $t$ -- , we cannot rule out, in principle, the 
possibility that the full metric in the bulk also is time dependent, and we have a nontrivial $j^i(r)$, but as we explained in the text,
this is not necessarily a problem.

\section{Conductivity using the Kubo formula at the horizon}

In this Appendix we give details about the calculation using the Kubo formula
\bea
\sigma(\omega)=-{\rm Im}\left(\frac{G_{yy}(\omega,\vec q)}{\omega}\right)\label{Kubo}
\eea
at the horizon.
The boundary term in the action is (we omit the scalar indices for now)
\bea
S_{yy} = -\frac{1}{2g_4^2}\int d^3x \sqrt{-g} X^{uy\mu y}A_y \partial_{\mu}A_y
\bigg|_{u\to 1} \qquad \mbox{(sum over $\mu$)}.
\eea
Note that in the massless case, i.e. for the pure $AdS_4$ black hole, the integrand above reduces to 
$\sqrt{-g} X^{uyuy}A_y \partial_u A_y$. In our case however, we have an extra term coming from the 
fact that $X^{uyty}$ does not vanish in the background \eqref{metricul}. Therefore, $S_{yy}$ is 
\bea
S_{yy} = -\frac{1}{2g_4^2}\int d^3x \sqrt{-g} \left(X^{uyuy}A_y \partial_{u}A_y + 
X^{uyty}A_y \partial_t A_y\right)\bigg|_{u\to 1}\label{Syy2}.
\eea
Since we are interested in the conductivity at $\omega=0$, one might naively think that we can simply 
ignore this second term above on the grounds that the time derivative will produce a term 
proportional to $\omega$ in the Green's function $G_{yy}$. However, the first term $\partial_u A_y$, 
when evaluated on-shell, will also be proportional to $\omega$, and in fact gives a (leading) finite contribution 
to the conductivity \eqref{Kubo}. Therefore, we can not simply ignore $\mathcal{O}(\omega)$ extra 
terms right from the beginning.

Thus, it seems that we need to go back to the Einstein's equations to obtain the back-reacted Weyl component $C^{uyty}$ at the horizon.  
However, we don't need to, since, in the general background \eqref{metricul}, we have 
\bea
X^{uyty}= -\frac{g_{ut}}{g_{tt}} X^{uyuy}.
\eea 
To simplify even more the integrand in \eqref{Syy2}, we can also take advantage of the infalling 
condition at the horizon (valid for $u\to 1$) given in \eqref{EFA} to write $\partial_t A_y$ in terms of 
$\partial_u A_y$. Thus, all in all we have
\bea
S_{yy} = -\frac{1}{2g_4^2}\int d^3x \sqrt{-g} X^{uyuy} A_y \partial_{u}A_y 
\left(1+\frac{u^2}{r_0 a(u) }\frac{g_{ut}}{g_{tt}}\right)\bigg|_{u\to 1}\label{Syy3}
\eea
where
\bea
a(u)= \frac{u^2}{r_0} \left(\frac{g_{ut}}{g_{tt}}+ \sqrt{\frac{g_{ut}^2}{g_{tt}^2}+\frac{g_{uu}}{g_{tt}}}\right).
\eea
The second term inside the parentheses in \eqref{Syy3} was not present in 
the massless case, when $g_{ut}$ is zero. However, expanding about $u=1$ gives
\bea
1+\frac{u^2}{r_0 a(u) }\frac{g_{ut}}{g_{tt}} = 1 + \mu \, \mathcal{O}(1-u)^{1/2}
\eea
which makes it irrelevant for the computation of $S_{yy}$ at the horizon. 

The Green's function we need is therefore
\bea
G_{yy}^{(A)} = -\frac{3r_0}{L^2g_4^2}\left(1+4\gamma+h_A(\mu^2)\right)(1-u)
\frac{\partial_u A_y^A(\omega)}{A_y^A(\omega)}\bigg|_{u\to 1}
\eea
where $h_A$ was defined in (\ref{hA}).
The last step is to obtain $A_y$ near the horizon. The equation of motion for $A_y$ is
\bea
A_y''+\alpha A_y' +\beta A_y=0
\eea
where
\bea
\alpha &\equiv& \frac{\partial_t \left(\sqrt{-g}X^{tyuy}\right)
+\partial_u \left(\sqrt{-g}X^{uyuy}\right)-2i\omega\sqrt{-g}X^{tyuy}}{\sqrt{-g}X^{uyuy}} \\
\beta &\equiv& -\frac{q^2 \sqrt{-g}X^{xyxy}+\omega^2 \sqrt{-g}X^{tyty}
+i\omega\left(\partial_t+\partial_u\right)(\sqrt{-g}X^{tyty})}
{\sqrt{-g}X^{uyuy}}.
\eea
Inserting the ansatz $A_y=(1-u)^b F(u)$ into the equation of motion we again 
obtain $b=\pm \frac{i L^2 \omega}{3r_0}$ near $u=1$ (independent of $\mu$!). 
Using the $+$ (infalling) value for $b$ again into the equation for $A_y$ near $u=1$, we get
\bea
\frac{F'(1)}{F(1)}\simeq i\omega \left[\frac{L^2(3-28\gamma)}{6r_0(1+4\gamma)}
+\mathcal{O}(\mu^2)\right] + \mathcal{O}(\omega^2)
\eea
for $q^2=0$ and small $\omega$. Therefore
\bea
\frac{\partial_u A_y(\omega)}{A_y(\omega)}\bigg|_{u\to 1} =  
i\omega\frac{L^2}{3r_0}\frac{1}{1-u}+i\omega\left(\frac{L^2(3-28\gamma)}{6r_0(1+4\gamma)}
+\mathcal{O}(\mu^2)\right) + \mathcal{O}(\omega^2).
\eea
Putting all this into the Kubo formula \eqref{Kubo} we obtain once again
\bea
\sigma^{(A)} = \frac{1}{g^2_4}\left(1+4\gamma+h_A(\mu^2)\right)
\eea
which matches the other results we obtained.

\bibliography{bhabjm}

\providecommand{\href}[2]{#2}\begingroup\raggedright\begin{thebibliography}{10}

\bibitem{Aharony:2008ug}
O.~Aharony, O.~Bergman, D.~L. Jafferis, and J.~Maldacena, ``{N=6 superconformal
  Chern-Simons-matter theories, M2-branes and their gravity duals},''
  \href{http://dx.doi.org/10.1088/1126-6708/2008/10/091}{{\em JHEP} {\bf 0810}
  (2008)  091},
\href{http://arxiv.org/abs/0806.1218}{{\tt arXiv:0806.1218 [hep-th]}}.

\bibitem{Sachdev:2011wg}
S.~Sachdev, ``{What can gauge-gravity duality teach us about condensed matter
  physics?},'' {\em Ann.Rev.Condensed Matter Phys.} {\bf 3} (2012)  9--33,
\href{http://arxiv.org/abs/1108.1197}{{\tt arXiv:1108.1197 [cond-mat.str-el]}}.

\bibitem{Myers:2010pk}
R.~C. Myers, S.~Sachdev, and A.~Singh, ``{Holographic Quantum Critical
  Transport without Self-Duality},''
  \href{http://dx.doi.org/10.1103/PhysRevD.83.066017}{{\em Phys.Rev.} {\bf D83}
  (2011)  066017},
\href{http://arxiv.org/abs/1010.0443}{{\tt arXiv:1010.0443 [hep-th]}}.

\bibitem{Gomis:2008vc}
J.~Gomis, D.~Rodriguez-Gomez, M.~Van~Raamsdonk, and H.~Verlinde, ``{A Massive
  Study of M2-brane Proposals},''
  \href{http://dx.doi.org/10.1088/1126-6708/2008/09/113}{{\em JHEP} {\bf 0809}
  (2008)  113},
\href{http://arxiv.org/abs/0807.1074}{{\tt arXiv:0807.1074 [hep-th]}}.

\bibitem{Terashima:2008sy}
S.~Terashima, ``{On M5-branes in N=6 Membrane Action},''
  \href{http://dx.doi.org/10.1088/1126-6708/2008/08/080}{{\em JHEP} {\bf 0808}
  (2008)  080},
\href{http://arxiv.org/abs/0807.0197}{{\tt arXiv:0807.0197 [hep-th]}}.

\bibitem{Iqbal:2008by}
N.~Iqbal and H.~Liu, ``{Universality of the hydrodynamic limit in AdS/CFT and
  the membrane paradigm},''
  \href{http://dx.doi.org/10.1103/PhysRevD.79.025023}{{\em Phys.Rev.} {\bf D79}
  (2009)  025023},
\href{http://arxiv.org/abs/0809.3808}{{\tt arXiv:0809.3808 [hep-th]}}.

\bibitem{Auzzi:2009es}
R.~Auzzi and S.~P. Kumar, ``{Non-Abelian Vortices at Weak and Strong Coupling
  in Mass Deformed ABJM Theory},''
  \href{http://dx.doi.org/10.1088/1126-6708/2009/10/071}{{\em JHEP} {\bf 0910}
  (2009)  071},
\href{http://arxiv.org/abs/0906.2366}{{\tt arXiv:0906.2366 [hep-th]}}.

\bibitem{Mohammed:2010eb}
A.~Mohammed, J.~Murugan, and H.~Nastase, ``{Looking for a Matrix model of
  ABJM},'' \href{http://dx.doi.org/10.1103/PhysRevD.82.086004}{{\em Phys.Rev.}
  {\bf D82} (2010)  086004},
\href{http://arxiv.org/abs/1003.2599}{{\tt arXiv:1003.2599 [hep-th]}}.

\bibitem{Lambert:2011eg}
N.~Lambert, H.~Nastase, and C.~Papageorgakis, ``{5D Yang-Mills instantons from
  ABJM Monopoles},'' \href{http://dx.doi.org/10.1103/PhysRevD.85.066002}{{\em
  Phys.Rev.} {\bf D85} (2012)  066002},
\href{http://arxiv.org/abs/1111.5619}{{\tt arXiv:1111.5619 [hep-th]}}.

\bibitem{Nastase:2000tu}
H.~Nastase and D.~Vaman, ``{On the nonlinear KK reductions on spheres of
  supergravity theories},''
  \href{http://dx.doi.org/10.1016/S0550-3213(00)00214-5}{{\em Nucl.Phys.} {\bf
  B583} (2000)  211--236},
\href{http://arxiv.org/abs/hep-th/0002028}{{\tt arXiv:hep-th/0002028
  [hep-th]}}.

\bibitem{Cvetic:1999xx}
M.~Cvetic, S.~Gubser, H.~Lu, and C.~Pope, ``{Symmetric potentials of gauged
  supergravities in diverse dimensions and Coulomb branch of gauge theories},''
  \href{http://dx.doi.org/10.1103/PhysRevD.62.086003}{{\em Phys.Rev.} {\bf D62}
  (2000)  086003},
\href{http://arxiv.org/abs/hep-th/9909121}{{\tt arXiv:hep-th/9909121
  [hep-th]}}.

\bibitem{Parikh:1997ma}
M.~Parikh and F.~Wilczek, ``{An Action for black hole membranes},''
  \href{http://dx.doi.org/10.1103/PhysRevD.58.064011}{{\em Phys.Rev.} {\bf D58}
  (1998)  064011},
\href{http://arxiv.org/abs/gr-qc/9712077}{{\tt arXiv:gr-qc/9712077 [gr-qc]}}.

\bibitem{Hartnoll:2011fn}
S.~A. Hartnoll, ``{Horizons, holography and condensed matter},''
\href{http://arxiv.org/abs/1106.4324}{{\tt arXiv:1106.4324 [hep-th]}}.

\end{thebibliography}\endgroup
\bibliographystyle{utphys}

\end{document}